\documentclass{physeauth}
\usepackage{graphicx}
\usepackage{amsmath}
\usepackage{amssymb}

\newcommand{\vet}{\boldsymbol}

\newcommand{\be}{\begin{equation}}
\newcommand{\bs}{\begin{split}}
\def\a{\alpha}

\def\G{\Gamma}

\def\D{\Delta}

\def\r{\rho}
\def\w{\omega}

\begin{document}

\begin{frontmatter}

\title{Shuttle instabilities: semiclassical phase analysis}

\author[]{A.~Donarini\thanksref{thank1}}
and
\author[]{A.-P.~Jauho}

\address{Mikroelektronik Centret, Building 345  east, Technical
University of Denmark, DK-2800 Lyngby, Denmark}


\thanks[thank1]{
Corresponding author. E-mail: ad@mic.dtu.dk}

\begin{abstract}

We present a semiclassical analysis of the instability  of an
electron shuttle composed of three quantum dots: two are fixed and
coupled via leads to electron resevoirs at $\mu_{L,R}$ with $\mu_L
\gg \mu_R$, while the central dot is mounted on a classical
harmonic oscillator. The semiclassical analysis, which is valid if
the central dot oscillation amplitude is larger than the quantum
mechanical zero point motion, can be used to gain additional
insight about the relationship of resonances and instabilities of
the device.

\end{abstract}

\end{frontmatter}

In nano-electromechanical systems (NEMS) the electrical and
mechanical properties are deeply interconnected. An archetypal
NEMS device consists of a movable object connected to leads
\cite{Gor-prl-98}. The charge distribution gives rise to an
electrical force, which influences the mechanical dynamics, while
the position of the movable object determines the tunneling rates
from the leads and thus influences the electrical dynamics of the
system. A current through the device can sustain mechanical
oscillations even in the presence of damping. An interesting
regime of transport arises when only one electron per cycle is
transferred from the left to the right lead. Due to the position
dependent tunneling amplitude the movable part gets charged when
near to the left lead, then the electrostatic force pushes it
towards the right lead where the now enhanced tunneling rate helps
the release of the electron.

We describe the electronic part with the density matrix formalism
and couple the master equation to a classical equation of motion
for the central dot position \cite{Gor-prl-98}. We perform a
linear instability analysis of the system to pinpoint where the
equilibrium solution for the system becomes unstable leading to
Hopf bifurcations; the relative phase between charge, velocity and
position of the unstable rotating solution shows the features of
shuttling in agreement with the quantum-phase-space description
\cite{Nov-prl-03}.

The semiclassical approach is justified by the quantum-classical
correspondence since the dot oscillations are bigger than the
minimum quantum amplitude. However the mean field approach for the
electric part neglects the effect of shot noise and the damping
factor threshold is usually very much reduced compared to the
quantum treatment. The semiclassical analysis is also much easier
to handle numerically.
%
\section{The model}
%
Consider a simple NEMS consisting a small metallic grain (or a QD)
connected to two leads. The grain is moving in a parabolic
potential. 
Two additional dots, one on each side of the oscillating
component, fix the energy of the incoming and outgoing electrons
\cite{Arm-prb-02}. The Hamiltonian of the system is
$H=H_{mech}+H_{el}$, where
\begin{equation}
 H=H_{mech}+H_{el}
\end{equation}
where \begin{equation} \begin{split}
 H_{mech}&=\frac{p^2}{2m}+\frac{1}{2}m\w^2x^2\\
 H_{el}(x)&=\sum_{i,j=L,C,R}|i \rangle \epsilon_{ij}(x)\langle j|
\end{split}
\end{equation}
We assume strong Coulomb blockade regime, thus the vectors
$|i\rangle, i=L,C,R$, together with the empty state $|0\rangle$,
span the entire single particle Hilbert space of the device. The
matrix elements are explicitly:
\begin{equation}
 \epsilon(x)=\left[\begin{array}{ccc}
   \frac{\D V}{2} & t_L(x) & 0 \\
   t_L(x) & -\frac{\D V}{2x_0}x & t_R(x) \\
   0 & t_R(x) & -\frac{\D V}{2}
 \end{array}\right]
\end{equation}
$\D V$, the {\it device bias}, is the difference between the
energy of the left and the right dot.
$x_0$ is half the distance between the two outer dots and
represents the maximum amplitude of the inner dot oscillation. The
three dots are electrically connected only via a tunneling
mechanism. The tunneling length is given by $1/\a$ and the
tunneling strength depends on the position $x$ of the inner grain,
\begin{equation} \begin{split}
 t_L(x)&=V_0e^{-\a(x_0+x)}\\
 t_R(x)&=V_0e^{-\a(x_0-x)}
\end{split}
\end{equation}
The electric part of the NEMS is described in  the density matrix
formalism. The Hilbert space has dimension four, though the
corresponding four-by-four density matrix can be reduced to an
effective three-by-three matrix since the elements of the form
$\rho_{0i}, i=L,C,R$ are decoupled from the others and $\rho_{00}$
can be eliminated using $\sum_i\rho_{ii}=1$. This effective
density matrix obeys the generalized master equation:
\begin{equation}\label{SSa}
 \dot{\rho}=-i[H,\rho]+ \Xi \rho
\end{equation}
The first term represents the coherent evolution of the electrons
in the three dots when isolated from the leads. The coupling to
the leads is introduced in the wide band approximation following
Gurvitz \cite{Gur-prb-96} and gives the second term of the
equation ($\Xi \rho$):
\begin{equation}
 \Xi \rho=\G\left[\begin{array}{ccc}
   1-\rho_{LL}-\rho_{CC}-\rho_{RR} & 0             & -\rho_{LR}/2 \\
   0                               & 0             & -\rho_{CR}/2 \\
   -\rho_{RL}/2                   & -\rho_{RC}/2 & -\rho_{RR}    \\
 \end{array}\right]
\end{equation}
where $\G$ is the injection rate to the leads.
The equation of motion for the central dot position has a
conservative part deduced from the (quantum-)averaged Hamiltonian
$\langle H \rangle = H_{mech}+{\rm Tr}(\rho H_{el}(x))$ and a
phenomenological damping term:
\begin{equation}\label{SSd}
 \begin{split}
 \dot{x}=&\frac{\partial \langle H \rangle}{\partial p}=\frac{p}{m}\\
 \dot{p}=&-\frac{\partial\langle H \rangle}{\partial x}-\gamma p=-m\w^2x+\frac{\D V}{2x_0}\rho_{CC} -\gamma p\\
 &+ \a t_L(\rho_{CL}+\rho_{LC})
  - \a t_R(\rho_{CR}+\rho_{RC})
\end{split}
\end{equation}
%
%
\section{Instability analysis}
\begin{figure}[h]
\begin{center}\leavevmode
\includegraphics[width=\linewidth, angle=0]{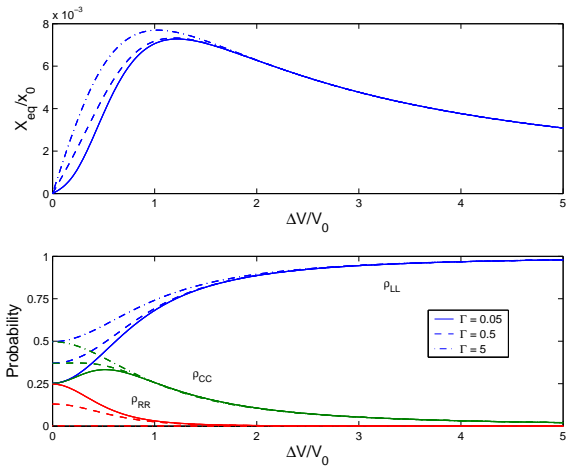}
\caption{Equilibrium position (a) and equilibrium electronic
configuration (b) as a function of the device bias for three
different injection rates $\Gamma=0.05,0.5,5$.}\label{equilsol}
\end{center}
\end{figure}

%
In the shuttling regime the electrical and the mechanical
component of the system evolve with the same time scale. Thus it
is important to treat all degrees of freedom on equal footing: we
introduce a set of generalized coordinates $q$, and write the Eqs.
(\ref{SSa},\ref{SSd}) as
\be \label{ODE}
 \dot{q}_i = F_i({\bf q})
\end{equation}
and linearize around the equilibrium,($0 \equiv {\bf F}({\bf q}^0)$): 
\be \label{linear} \dot{q}_i = {\mathcal M}_{ij} (q_j-q^0_j) + ...
\end{equation}
where ${\mathcal M}_{ij}=\partial_jF_i|_{\vet{q}=\vet{q}^0}$.
In general, even if the velocity of the central dot is zero, a
current can flow through the system and is given by $I=\Gamma
\rho_{RR}$. Let us study the equilibrium position and electronic
configuration of the three dot system as a function of the device
bias. At $\Delta V=0$ the central dot is at rest in the middle (we
will take this as zero position). This unique solution appears
natural, but in fact is not obvious if one recognizes the presence
of an exchange force that acts on the dot (i.e. the last line of
Eq. \ref{SSd})\footnote{We thank the Chalmers group for suggesting
this point during private communications.}. As the difference
between the energy of the left and right dot is increased the
equilibrium position of the central dot shifts to the right, but
then returns to zero again for high device biases
(Fig.\ref{equilsol}a). This reveals a competition between the
electrostatic force
and the progressively dominating decoupling of the three dots due
to the mismatch of their energy levels. The equilibrium occupation
of the three dots also reflects this decoupling
(Fig.\ref{equilsol}b). In the limit of high device biases the left
dot has the highest probability of being occupied while the others
are almost empty. At small biases the equilibrium occupation is
strongly dependent on the injection rate $\Gamma$. In this regime
the ratio between the injecting rate and the bare tunneling
probability $r=\Gamma/(V_0e^{-\alpha x_0})$ accurately describes
the scenario. If $r\ll 1$ the three dots are almost isolated and
their coherent oscillations are damped by the contacts, but so
slowly that the three dots tend to a uniform stationary
occupation. In the opposite limit ($r\gg 1$) the left dot is
continuously refilled and the right emptied at a rate which
overcomes the dot's dynamical response. The left and central dot
share the occupation probability while the right dot is empty and
the empty state occupation probability vanishes.

\begin{figure}[h]
\begin{center}\leavevmode
\includegraphics[width=\linewidth]{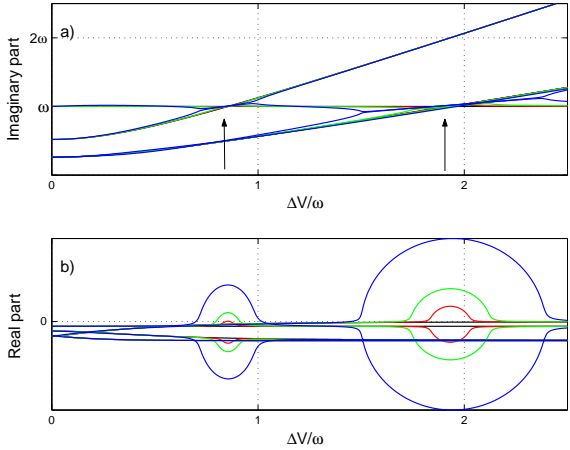}
\caption{Imaginary (a) and real (b) part of the spectrum of
${\mathcal M}$ for different values of tunneling length.
$\alpha=0, 0.1,0.2,0.5$. The instability grows with $\alpha$. Only
the four eigenvalues with positive imaginary part are plotted. The
system is symmetric to complex conjugation. The three eigenvalues
with zero imaginary part are omitted for simplicity: they
represent always damped modes. }\label{spectra}
\end{center}
\end{figure}

The quantum description of the shuttle device shows that the
central dot can oscillate with an amplitude which is much bigger
than the minimum uncertainty length \cite{Nov-prl-03}. We expect
the equilibrium solution of the semiclassical description to be
unstable in the same parameter range (except for the damping: the
lack of noise sources in the present semiclassical approach
reduces the instability threshold). The vector ${\bf q}=
[\;\r_{LL} \; \r_{CC} \; \r_{RR}\; \r_{LC}\; \r_{CL}\; \r_{LR}\;
\r_{RL}\;
 \r_{CR}\; \r_{RC}\; x\; p\;]$
has an \emph{electrical} part (the first 9 components) and a
\emph{mechanical} part (the last 2 components).
The matrix $\mathcal M$ is written in a block form, consistent
with this natural separation: \be
 {\mathcal M}=\left[
\begin{array}{cc}
 {\mathcal A} & {\mathcal C}\\
 {\mathcal D} & {\mathcal B}\\
\end{array}\right]
\end{equation}
The eigenvalues of  ${\mathcal M}$ represent the characteristic
frequencies and damping for small oscillation of the position and
the density matrix around the equilibrium configuration. If at
least one of the eigenvalues of ${\mathcal M}$ has a positive real
part then the equilibrium solution is unstable. In the general
case all the four blocks of the matrix ${\mathcal M}$ depend on
both the electrical and mechanical degree of freedom. But if we
set $x_0=1/\alpha$ and take the limit $\a \to 0$, the off-diagonal
blocks of the $\mathcal M$ (${\mathcal C}$ and ${\mathcal D}$)
vanish and the eigenvalues are those of ${\mathcal A}$
(electrical) and ${\mathcal B}$ (mechanical). The imaginary part
of the spectrum consists of the Bohr frequencies (i.e. all
possible differences between the eigenvalues) of the isolated
electron system and the mechanical frequency of the oscillator
(Fig.\ref{spectra}a). The real part of the spectrum, though, is
negative due to the presence of the leads and the mechanical
damping (Fig.\ref{spectra}b). As $\alpha$ is increased, the
spectrum of ${\mathcal M}$ gets a positive real part: two
instability regions appear (Fig.\ref{spectra}b). They have the
form of Hopf bifurcations: for these parameters  the solution of
Eq.\ref{ODE} is a limit cycle. Correspondingly, in the unstable
regime, the imaginary part of the spectrum shows crossings between
the ``electrical'' and ``mechanical'' frequencies (arrows in
Fig.\ref{spectra}a). These are resonances between the oscillator
and the electrical modes of the three dots. They appear $\Delta
V\simeq$ $\omega$ and $2\omega$. The slight shift from the exact
values is due to the bare interaction $V_0$ which modifies the
electronic spectrum of the Hamiltonian from the decoupled $0, \pm
\Delta V/2$ (which remains as asymptotic behavior).

\begin{figure}[h]
\begin{center}\leavevmode
\includegraphics[width=\linewidth]{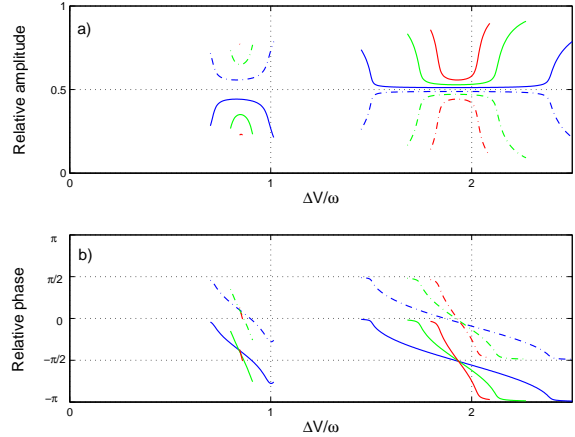}
\caption{Eigenvector analysis as a function of device bias. a) The
relative amplitude of the electrical (full) and mechanical (dot
dashed) components of the unstable eigenvector. The relative
amplitude of the two components tends to be equal in the
instability region and the effect is more prominent for higher
 $\alpha$. b) The phase of position (full) and velocity (dot dashed)
with respect to the phase of the charge in the central dot for the
unstable eigenvector.}\label{ampliphase}
\end{center}
\end{figure}

Only an analysis of the eigenvectors can demonstrate the actual
oscillation of both the electric and mechanical components in the
limit cycle solution. For this reason we have studied the
electrical and mechanical weight of the eigenvectors in the
unstable regime (Fig.\ref{ampliphase}a). The larger the real part
of the unstable eigenvalue, the higher is the mixing of the two
components.

In the quantum description
\cite{Nov-prl-03}, the shuttling instability is characterized by a
strong correlation between charge, position and velocity of the
oscillating dot. Therefore we study next  the relative phase
between charge, velocity and position of the unstable solution,
even if a priori is it not clear whether this property will be
maintained in the limit cycle stationary solution. The phase
analysis shows, for a given charge configuration (that for
definiteness we take maximal) a decrease in the position and
velocity phase: for both resonances the charged dot passes from
maximal position and zero velocity to minimal position and zero
velocity. The range of phase rotation is decreasing together with
the instability in the case with $\Delta V \approx \omega$
(Fig.\ref{ampliphase}b). This rotation of the charged dot
configuration in phase space is also in qualitative agreement with
the quantum analysis.

%
\section{Conclusions}
In this work we have analyzed the instabilities of a triple-dot
shuttle using a semiclassical approach, motivated by the
``semiclassical behavior'' of the quantum description. We were
able in this simple and intuitive framework to identify the
correspondence between instability and resonance in the NEMS. Also
the rotation of the phase space configuration of the charged dot
in the shuttling regime was detected. More work is needed to
understand whether other characteristics of the system such the
higher order instabilities ($\Delta V \approx 3\omega, 4\omega
,... $) are pure quantum phenomena.

\end{document}